# Liquid crystalline behaviour of self-assembled Laponite/PLL-PEG nanocomposites

Peicheng Xu, Zhongyang Xing, Erika Eiser*

*Cavendish Laboratory, University of Cambridge, Cambridge CB3 0HE, United Kingdom*

Yang Lan

*Chemistry Department, University of Cambridge, Cambridge CB2 1EW, United Kingdom*

## Abstract

Synthetic Laponite-clay particles with platelet-like shape display strong aging when dispersed in aqueous solutions, preventing latter from reaching their natural liquid-crystalline equilibrium state. Here we introduce a facile method that successfully prevents this aging behaviour and enables accessing the systems' liquid-crystal and crystalline phases. We graft the comb-like polymer, poly(L-lysine)-g-poly(ethylene glycol) (PLL-PEG), onto the clay surfaces from solution, thereby screening the negative surface charges and thus ensuring steric stabilisation. We show zeta-sizer and rheology measurements, respectively, confirming complete steric coating and that aging of dilute samples is completely suppressed even after a year. Using evaporation as means to concentrate the particles, we observe various liquid crystalline textures under a polarized optical microscope (POM). Upon sequential spreading and drying, we are also able to obtain transparent films with hierarchical architecture.

## Introduction

The synthetic clay, Laponite, has been studied extensively because of its unusually strong aging behaviour when suspended in water. Already very small weight fractions of Laponite can render aqueous suspensions either a gel or glassy state, depending on the overall ionic strength of the suspensions – a property that is utilized in industrial applications ranging from cosmetics to additives in drilling fluids.[1-4] This property is attributed to the peculiar charging state of the platelets in water, which is due to their two-dimensional crystalline architecture exposing negative charges on the flat surfaces while the rims are positively charged (Fig. 1a).[5,6] The resulting complex interaction potential between two disks in water is responsible for the aging and eventual kinetic arrest of the samples. But it also limits the platelets' solubility in water to a few weight percent, which is way below the isotropic-to-nematic transition of these systems. However, there has been great interest in accessing the liquid-crystalline (LC) properties of these disk-like colloids to test predictions of various columnar and discotic smectic phases.[7,8]

Only few colloids with disk-like shape exist while the majority of natural and synthetic colloidal systems are rod-like. Hence, clays represent a natural candidate for studying their various LC phases. Indeed, other clays like beidellite,[9] gibbsite,[10] montmorillite[11] and nontronite[12] form more easily liquid crystalline phases, all of them showing predominantly discotic-nematic phases, also because they often have only one type of charge in water (typically negative) and their larger aspect ratio. Laponite, however, has resisted so far to various efforts in sterically stabilizing them such that high enough concentrations to the phase transition to a nematic state could be reached. Hansen et al. have performed a number of studies in which Laponite solutions were evaporated in a cuvette.[13] Such slowly concentrated samples showed indeed a nematic appearance under crossed polarizers, however, this may have been due to the planar alignment of the disks in the vicinity of the flat container walls, while the interior remained in a disordered arrested state (Fig. 1b).

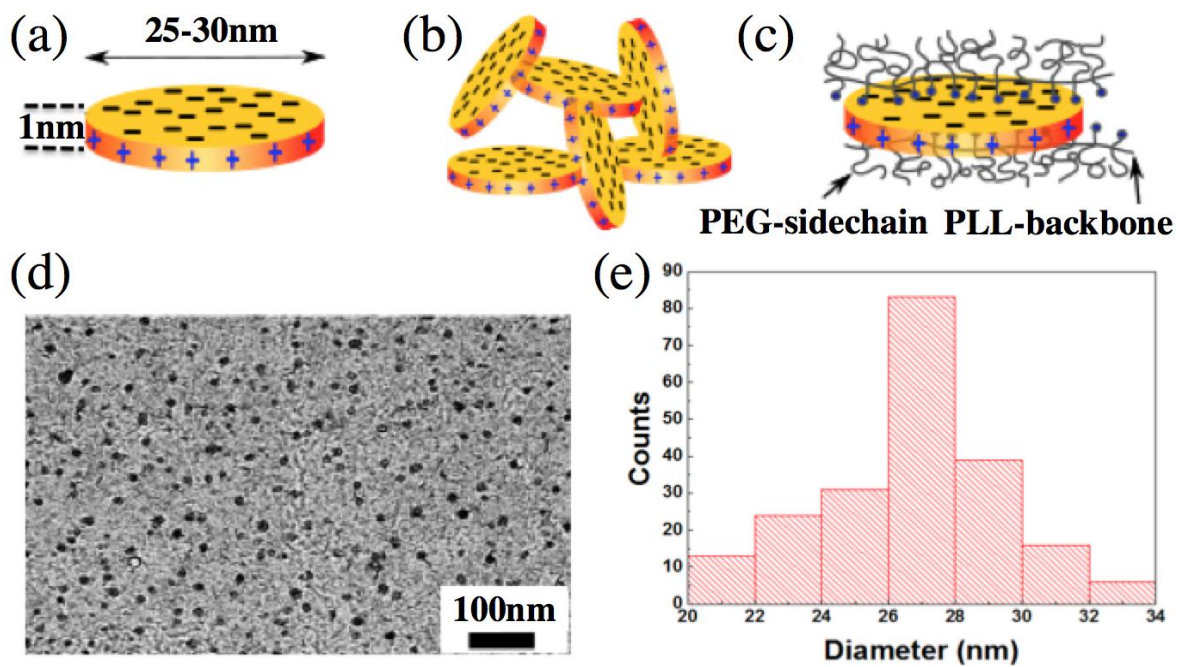

**Figure 1: (a)** Cartoon of a single Laponite disk. When dispersed in water Laponite disks have a positively charged rim and negatively charged flat surfaces. **(b)** Schematic representation of a gel phase. **(c)** Illustration of a PLL-PEG coated Laponite platelet. Positive ammonium ions along the PLL-backbone bind to the negative charges on the Laponite surfaces while the PEG chains protrude into the aqueous phase. **(d)** TEM image of such sterically-stabilized Laponite particles and (e) a corresponding particle-diameter histogram counting about 200 particles.

Here we present first studies of sterically-stabilized Laponite dispersions in which we were able to completely suppress the aging process, allowing us to study the isotropic-to-nematic phase transition. We start with the steric stabilization of Laponite particles in water using a comb-like polymer and the characterization of the suspensions with different polymer-coating strengths. Subsequently, we present drying droplet measurements demonstrating the successful transition from an isotropic clay suspension to a liquid crystalline texture. We further show unexpected phase-separation behaviour in the case of over-coating, in which the unbound polymers form droplets that undergo a LC transition and which subsequently serve as nucleation points for the LC-formation of the Laponite-rich phase. Finally, we demonstrate that these composite materials can be developed into thin, transparent films.

# Results and discussion

## Steric stabilization

First we demonstrate that steric stabilization of Laponite particles can prevent their aqueous suspensions from aging. The coating of the clay particles with PLL-PEG was tested by measuring the effective zeta potential of aqueous suspensions containing 0.5w%, 1.0w% or 2.0w% of un-coated particles respectively, and 1.0w% Laponite suspensions with three differently strong PLL-PEG coatings (Fig. 2). All samples containing the un-coated particles showed negative effective zeta potentials with nearly the same value around -35mV, representing an average between the positive rim and negative charges on the flat surfaces. These measurements show that the particles' effective zeta potential is independent of the particle concentrations in the range of concentrations used. Moreover, pH measurements (not shown here) also indicate that all suspensions reach an equilibrium value of around 8 after 1 day, which has been previously observed.

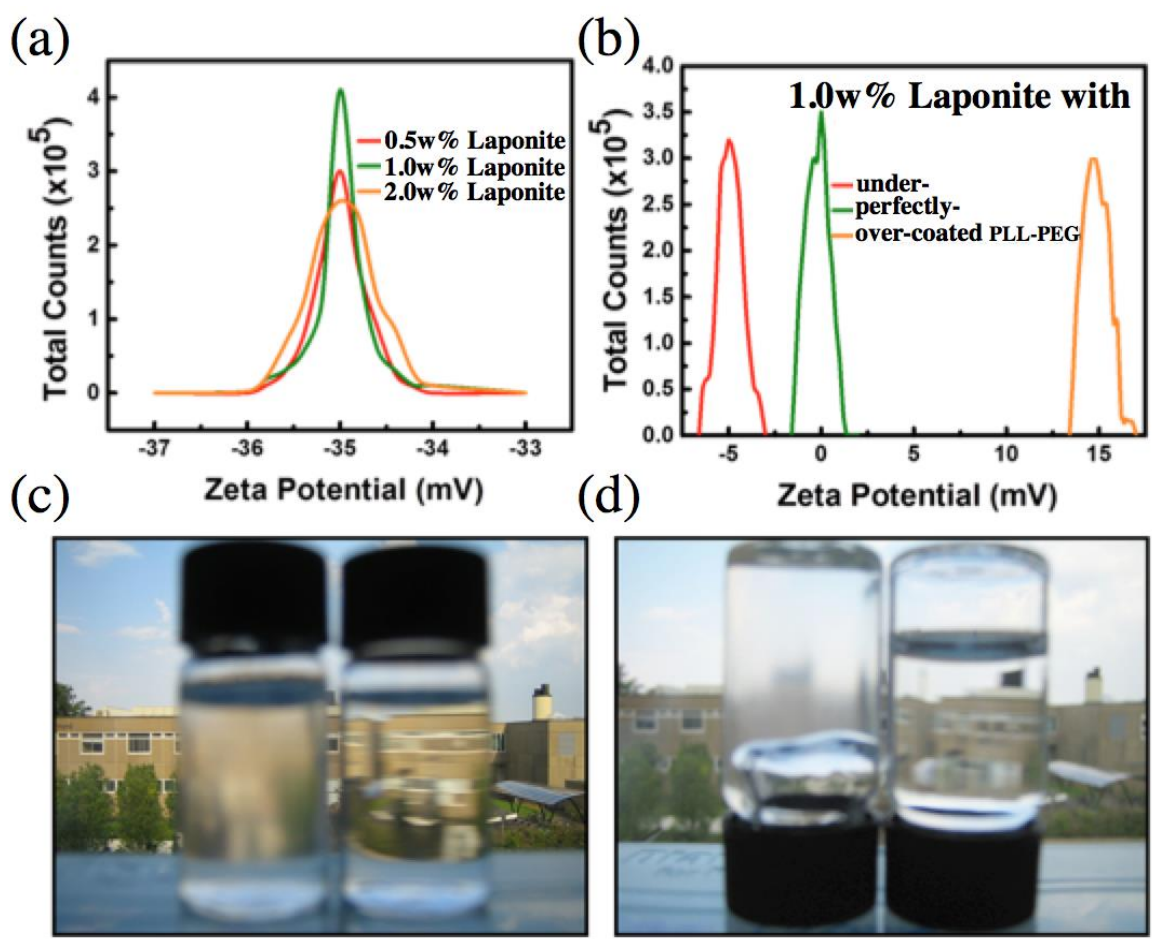

**Figure 2:** Zeta potential measurements for **(a)** pure Laponite solutions with a concentration of 0.5w%, 1.0w% and 2.0w% in deionized water and **(b)** for 50μL of 1.0w% Laponite solution mixed with 650μL (red line), 1200μL (green line) and 2400μL (orange line) PLL-PEG solutions respectively. The peak of the intermediate PLL-PEG concentration (green line) is located at -1.69mv, indicating negative surfaces charges have been neutralized. **(c)** Photograph of one-year old 1.0w% samples. The pure Laponite solution without coating was slightly turbid (*left*) while the PLL-PEG-coated Laponite solution still looked fully transparent. **(d)** The same samples as in **(c)** but turned upside down. While the pure Laponite solution had clearly become a gel the PLL-PEG-coated one was still completely liquid and flowed easily.

In order to minimize the negative charges on the clay surfaces we grafted PLL-PEG chains to them from solution, using the fact that the PLL backbone is positively charged. The radius of gyration $R_g$ of the 5kDa PEG side-chains in water is about 2.82 nm,[14] while the distance between two neighbouring grafting points of the PEG chains along the PLL backbone is only 1.18nm.[15] Therefore the PLL-PEG chains will take the shape of a stretched bottle-brush. When added to Laponite solutions the charged PLL backbone adsorbs to the negatively charged Laponite surface in a flat and stretched configuration with the PEG forming a sterically stabilizing brush due to their close spacing (see cartoon in Fig. 1c).[16,17] The length of the stretched PLL backbone is around 35 nm, similar to the average diameter of a Laponite disk. However, many configurations are possible including the possibility that a single PLL-PEG chain can link two particles. In order to avoid initial bridging between particles we prepared first relatively dilute Laponite suspensions to which we then added the polymer chains. In Fig. 2b we show the zeta-potential measurements of 1.0w% coated-Laponite dispersions for three different added PLL-PEG concentrations. For 50µL of 1w% Laponite suspensions mixed with 1200µL PLL-PEG we obtained an effective zeta potential of about -1.69mV, close to neutral, while samples with only 650µL added PLL-PEG showed a net negative and 2400µL added PLL-PEG a net positive surface charge. In the following we refer to these three coating situations as perfectly, under and over-coated samples. Long-term measurements made with samples containing pure Laponite particles and such with different coating strengths demonstrate that in suspensions with neutralized zeta potential the aging process could be completely suppressed due to the steric stabilization provided by the PEG side chains (Fig. 2c and d). Indeed, one-year-old 1.0w% samples containing perfectly-coated Laponite particles remained completely fluid, as can be seen when their container was turned upside down. In contrast, samples containing the un-coated particles were completely gelled, as observed previously.[2,18] Also, samples with the coated particles remained completely clear while that of the un-coated ones remained slightly turbid. Note that for higher PLL-PEG concentrations unbound chains will remain in solution, influencing the liquid crystalline behaviour of the drying solutions as will be shown later.

Micro-rheology measurements[19] further support our finding that the PLL-PEG coating can successfully suppress aging of the samples. In Fig. 3 we show the measured loss modulus for a freshly prepared 1.0w% solution with un-coated, and 1 year old samples with under-, perfectly- and over-coated particles. The loss modulus" from a freshly prepared, un-coated sample gave a viscosity of about 3mPas close to that of water, while 1-year-old samples with the three differently coated particles displayed a viscosity of about 2mPas. Note that we were not able to measure the 1-year-old pure Laponite sample as it was fully gelled and transferring into a capillary would have led to rejuvenation (fluidization) effects.[20,21] Further, no elastic contribution G' was detectable.

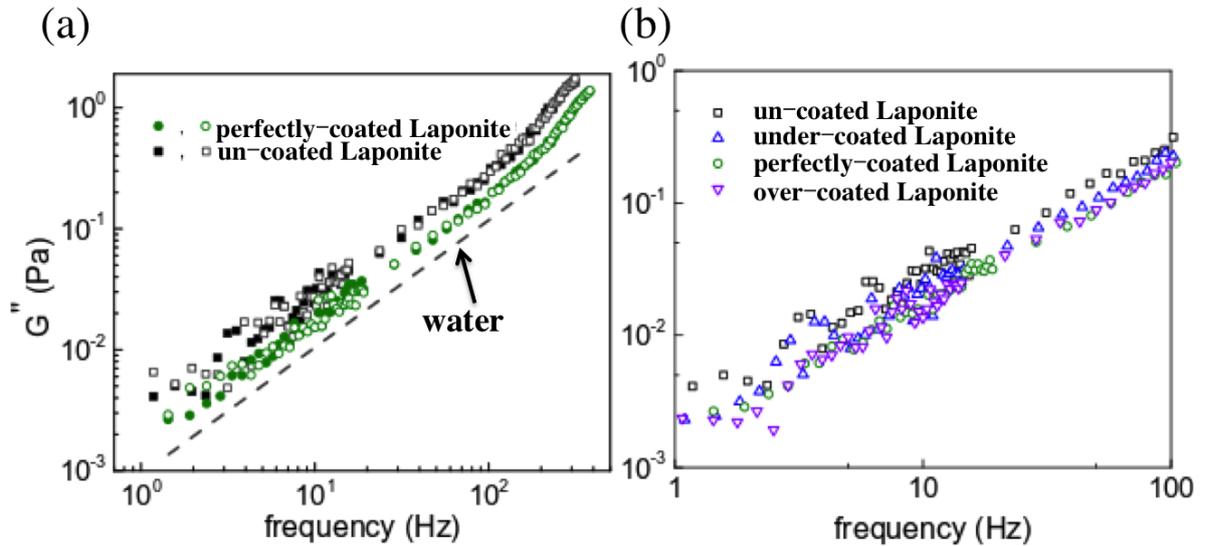

**Figure 3:** Micro-rheology measurements of the loss modulus $G''(\omega)$. The values extracted from the thermal fluctuations of a probe particle were similar in *x* and *y*-direction (closed & open symbols). **(a)** Comparing the 1.0w% freshly prepared, un-coated Laponite samples with a 1-year-old sample containing perfectly-coated samples: The extracted viscosities were 3 and 2 mPas, respectively. $G''(\omega)$ of water was plotted for reference. **(b)** For comparison we show $G''(\omega)$ measured for the all 4 preparations (as in Fig. 2).

**Liquid crystal formation**

We used polarized optical microscopy (POM; CPOM when polarizer and analyzer are crossed) to monitor the development of liquid crystalline (LC) textures in freshly prepared, drying droplet solutions containing the differently (un-)coated Laponite particles. For this we deposited a 120µL large droplet on a microscopy slide that was either sonicated in ethanol and then dried with nitrogen or plasma cleaned. We then placed the slides in a microscope with crossed polarizers and monitored the drying process. In all cases we observed that the droplet shrank without leaving a so-called coffee ring[22] behind – hence the particle concentration inside the drop increased steadily (Fig. 4). Pure Laponite solutions showed no birefringence throughout the entire drying process. The remaining solid material formed a uniform, transparent film with no birefringence.

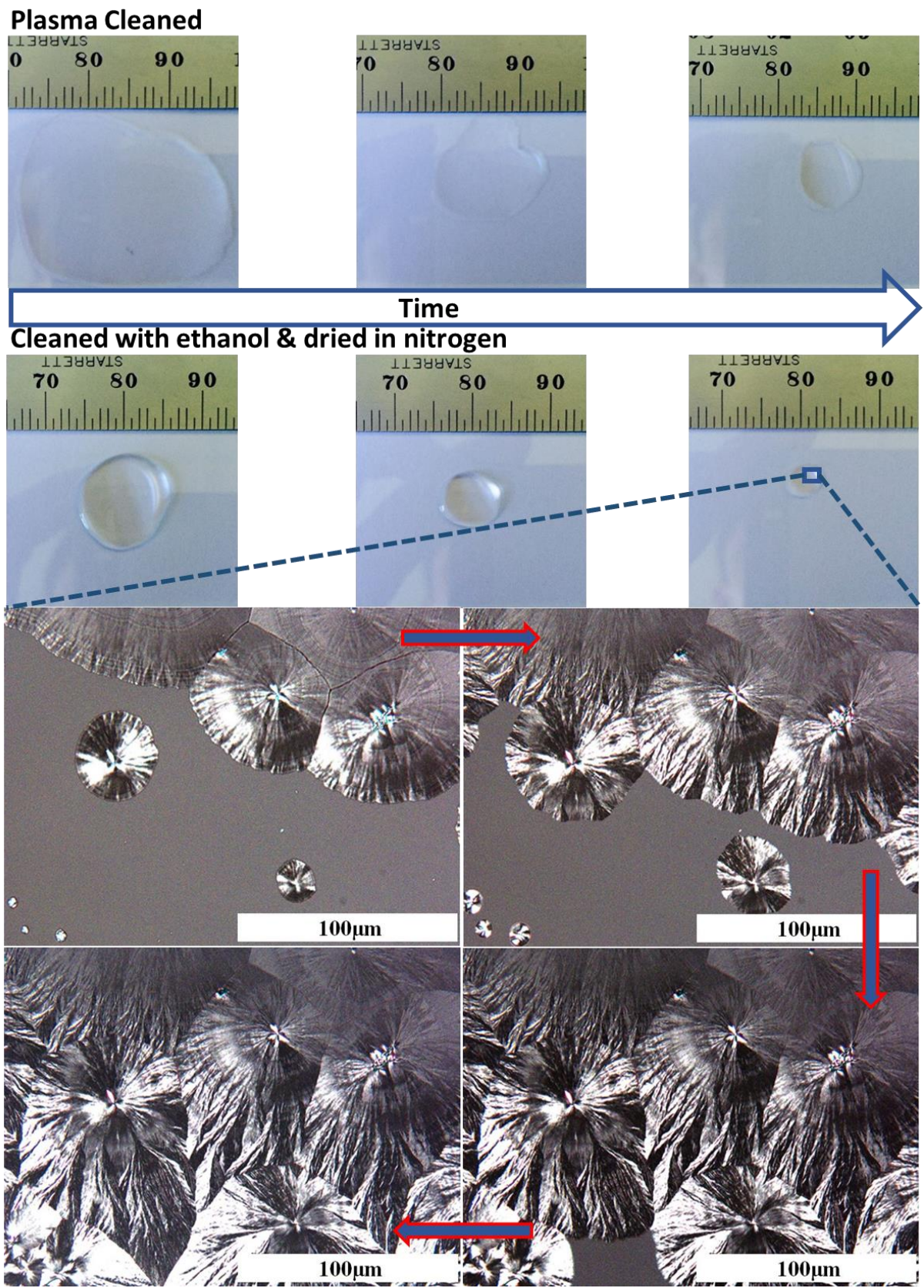

**Figure 4:** (*Top*) Photographs of drying droplets of over-coated Laponite particles with an initial particle concentration of about 1.0w% on two differently cleaned microscopy-glass slides. (Bottom) Microscope images taken with crossed polarizers at the rim of a drying droplet of an over-coated Laponite solution, starting at the onset of first visible LC textures. The blue arrows indicate the arrow of time.

Samples containing initially the differently coated 1.0w% Laponite particles on the other hand showed a sudden appearance of islands with liquid crystalline textures when the droplets shrank to about 3-5µL. As shown in Fig. 4 the entire sample became liquid crystalline eventually. The onset of a isotropic-to-nematic transition coincided roughly with volume fractions $\Phi$ obtained in Monte Carlo simulations for cut, hard spheres with an aspect ratio $l/d$ = 0.04, corresponding to Laponite disks with average diameter $d$ = 25nm and thickness $l$ = 1nm. We estimated this transition to occur at $\Phi_{low} \approx$ 12.1%, the onset of the 2-phase region, to $\Phi_{high} \approx$ 12.7% where the entire sample has been converted to an LC phase. These values were obtained assuming an average particle volume $V_p = \pi d^2 l/4$ and $\Phi = \pi d^2 l n/4$ or $\Phi = nd^3(\pi/4)(l/d)$ with $n$ being the particle's number density. For $l/d$ = 0.04, simulations predict a phase transition to occur at $nd^3$ = 3.85 as lower, and $nd^3$ = 4.04 as upper bound.[7,23] Because of the small sample volumes used we were unable to determine the exact concentration of the samples as they further dried. However, only small further changes of the LC textures were observed in the fully dried samples. In Fig. 5 we compare the almost dried textures observed in POM for the under-, perfectly, and over-coated samples as well as pure PLL-PEG solutions. All of them showed a perfectly transparent film, however under crossed polarizers they displayed different textures. Latter did not differ much from the completely dried films.

From the images in Fig. 5 we can make two important observations: One is that all different PLL-PEG coatings lead to the formation of liquid crystalline textures when the solutions are reaching the theoretical value for the isotropic-to-nematic transition. Secondly, also solutions of only PLL-PEG in pure water also show a LC-transition upon reaching a critical concentration. These observations raise the question whether the LC textures observed for the coated Laponite solutions are simply due to the PLL-PEG coatings alone or whether it is really given by the discotic nematic, columnar or smectic alignment of the platelets.

Both, textures for the under- and perfectly-coated samples show dendritic textures with predominantly dark and bright regions arranged in a bow-tie shape. Comparison with other systems displaying comparable symmetries we hypothesize that the under- and perfectly-coated Laponite solutions form columnar structures,[24] displaying s = ±1/2 defects.[25] Of course one could also assume a smectic structure with the discs either laying flat of tilted in the smectic layers,[26] which we can only really determine in small angle x-ray scattering (SAXS) experiments. The polydispersity along the long axis of the clay disks would in principle favor a lamellar DL (smectic) phase. A third possibility is that our under- and perfectly- coated samples may be in the fully crystalline phase. Fast forming crystals such as ice and rheo-casted Al-Si also show such dendritic growth as well.[27] The PLL-PEG with the PEG chains stretching out perpendicular to the clay disks has a further effect on the overall texture as the PEG chains themselves can crystallize upon drying, forming an additional anisotropic order and thus adding to the rather complex CPOM images. Again, to fully elucidate these structures we need to perform small angle x-ray scattering experiments in the future. Finally, we also note that the under-coated samples show consistently the same dendritic texture, however, with apparently smaller feature sizes.

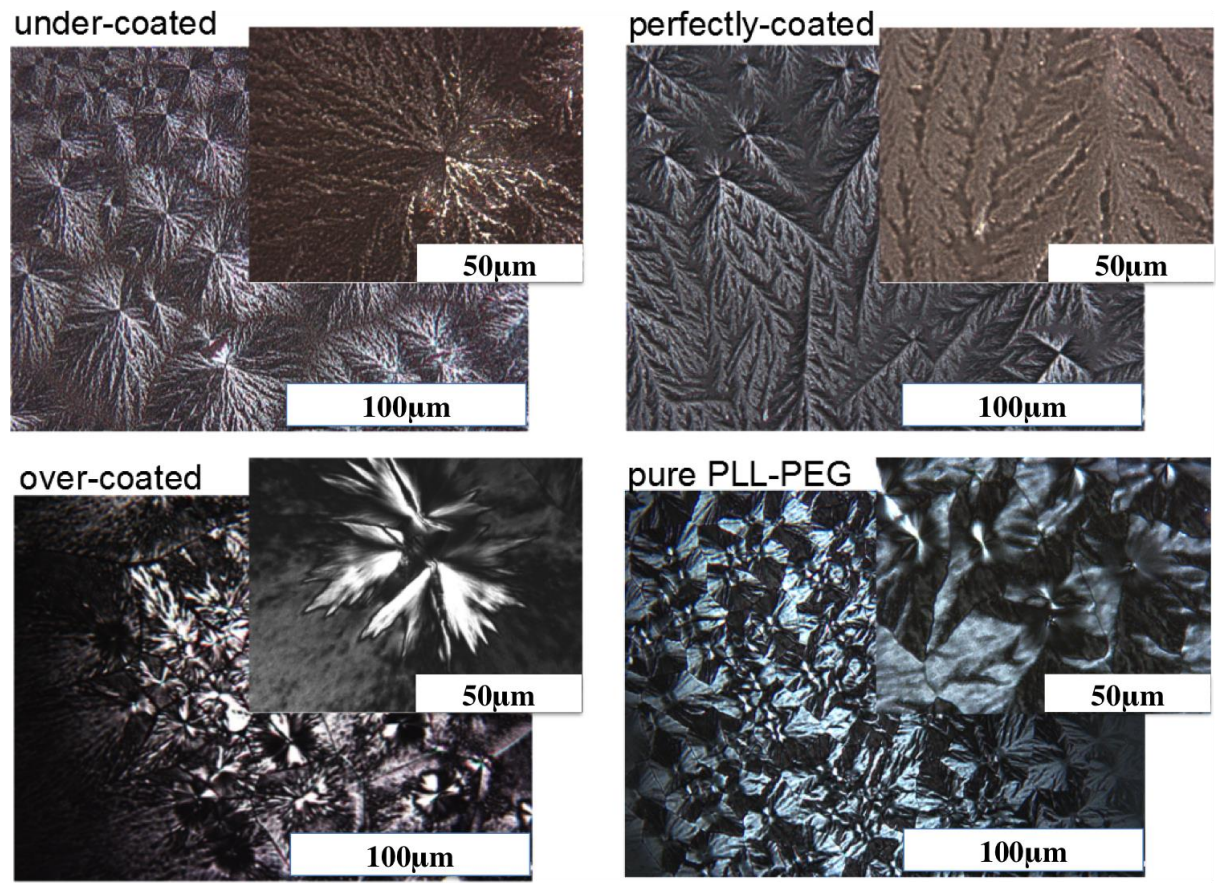

**Figure 5:** Microscopy images of almost dried droplet-films containing the differently coated Laponite particles and of a droplet prepared with only PLL-PEG. All images were taken with crossed polarizers. The larger images were taken with 20× and the insets with 100× magnification, with the scale bars being 100μm and 10μm respectively.

Studying the CPOM images of pure PLL-PEG solutions (Fig. 5,6) we see a very different liquid crystalline texture. Because of the close inter-anchor spacing between the grafted PEG chains, which is smaller than the radius of gyration of the corresponding free PEG chains, we can assume that the chains have a shape of a cylindrical bottle-brush with an approximate aspect ratio of 7. Hence, at high enough concentrations their solutions will undergo an isotropic-to-nematic transition, which we clearly observe in the drying-drop experiments. While we do not observe a classical Schlieren texture,[25] typical for a nematic sample sandwiched between two flat glass slides, we do observe a broken fan-like structure surrounded by more evenly changing areas, while different focal-point regions are separated by sharp, dark lines, which we interpret as disclination lines.

The over-coated Laponite samples show CPOM images that seem to combine both, the pure PLL-PEG and the perfectly coated samples (Fig. 5,6). This is visible through the large areas with a Maltese-cross texture, often observed in nematic samples, and the small region at the center of these crosses displaying a fan-like structure similar to those observed in the pure PLL-PEG solutions (Fig. 5). Such a behavior can easily be understood in terms of a phase-separation into a Laponite-rich and a PLL-PEG-rich region. Water is a good solvent for the PEG chains even at high PEG-concentrations,[28] hence we can expect a shape-induced phase separation between the sterically stabilized clay disks and the rod like PLL-PEG bottle-brushes, on top of the isotropic-to-nematic phase transition. By forming small PLL-PEG rich droplets, a clay-rich majority phase with nematic texture can evolve around these droplets. That we obtain a quasi-dual nematic texture is supported by the CPOM images in Fig. 6, where we rotate the sample while keeping polarizer and analyzer crossed at 90°. As we rotate the sample we see hardly any change in the overall texture.

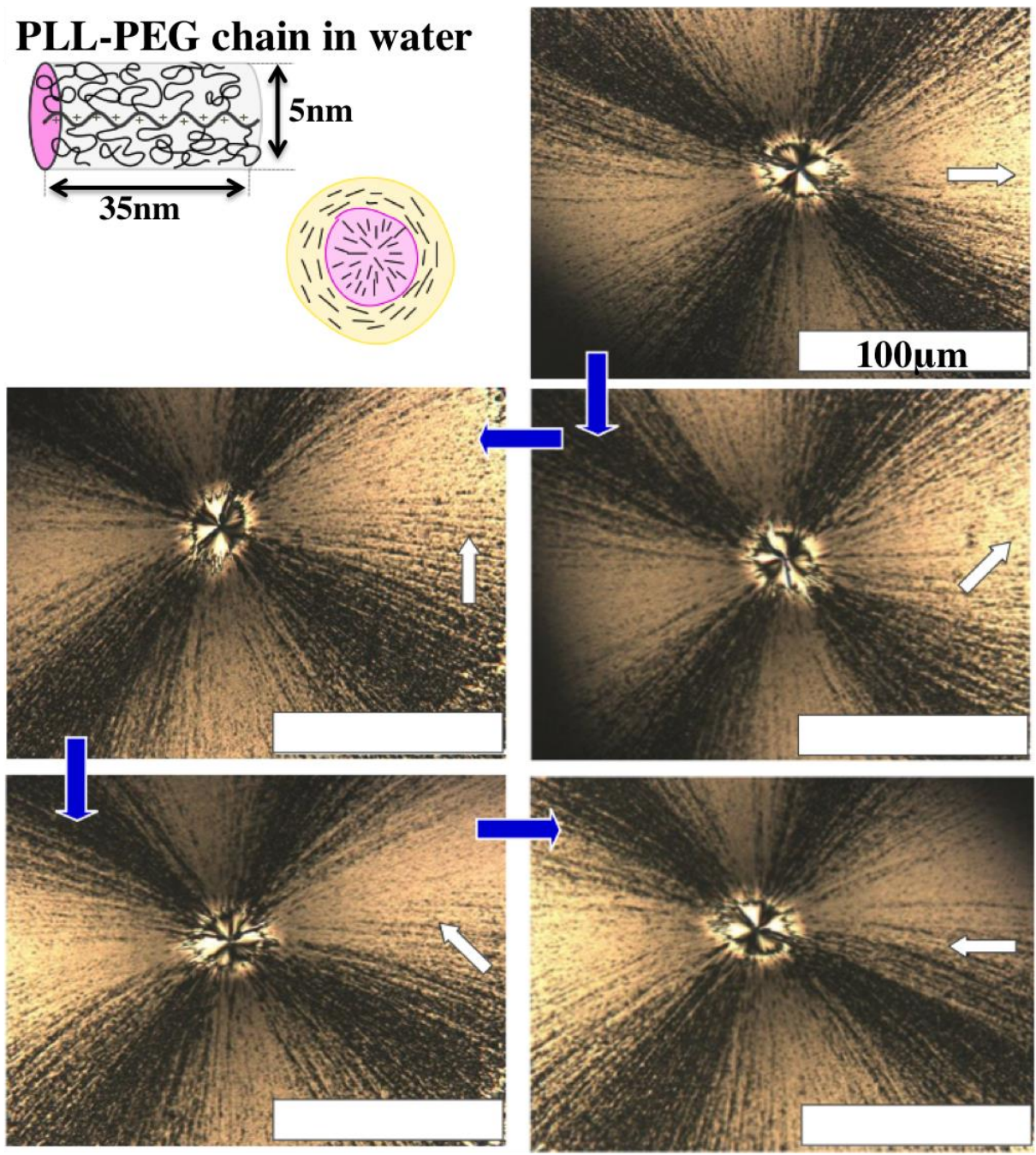

**Figure 6:** Alignment of PLL-PEG chains and cross-polarized images of drying droplets containing over-coated Laponite suspensions. Top left shows a cartoon of the cylindrical bottle-brush appearance of a PLL-PEG chain in water and a sketch of the alignment of these chains inside a droplet surrounded by the PLL-PEG coated Laponite clays forming concentric smectic layers. Microscope images were taken of these over-coated almost dry samples. The crossed polarizers were rotated (orientation indicated by the white arrow) by 180°.

**Layer-by-layer film preparation**

In addition to accessing the LC properties of clays, we wanted to utilize the smectic layer formation of polymer-coated clays to prepare a new type of coatings or films with mechanical properties similar to nacre. For this we prepared a larger suspension of over-coated Laponite particles. We placed a droplet of this suspension on a cleaned microscope cover slip, spread the solution with scalpel thinly and left it to dry. Subsequently, we placed another drop on the dried film, spread it and again left it to dry. We repeated this procedure 5 times leading to a transparent nano-composite film with a thickness of 2.6μm (Fig. 7). After breaking the glass slide with the film carefully we were able to study the fracture with a Scanning Electron Microscope (SEM): The top surface appears very smooth while the cross section of the film displays a multilayered structure with apparently sharp and more soft-edged thinner layers resembling a mille-feuille pastry. Malwitz et al., demonstrated that Laponite disks align when spread with a low shear rate and will quickly restore to isotropic arrangement when the shearing force ceases.[29] We argue that the sterically stabilizing PLL-PEG coating acts as the mortar between the drying 'clay-bricks'. Under the condition of excessive polymers in the bulk, the chain length of one PLL-PEG is long enough to interconnect two clay platelets. This bridging effect will further result in the observed layered organic-inorganic network. This is reminiscent of the structure in the natural mother of pearl or nacre that forms the iridescent material of sea shells.[30] These materials retrieve their mechanical toughness and resilience against breaking from brittle, uniformly sized aragonite crystals of $CaO_2$ that are held together by the more plastically deformable proteins. However, in our system, the inorganic crystals are 3 orders of magnitude thinner and smaller along the long axis than the calcium-oxide crystals in nacre. Therefor we do not achieve the same iridescent interplay with visible light. Nevertheless, our system represents simple, sustainable materials that could be developed for transparent, mechanically tough coatings that could find application for coatings of solar cells and other sensitive surfaces.

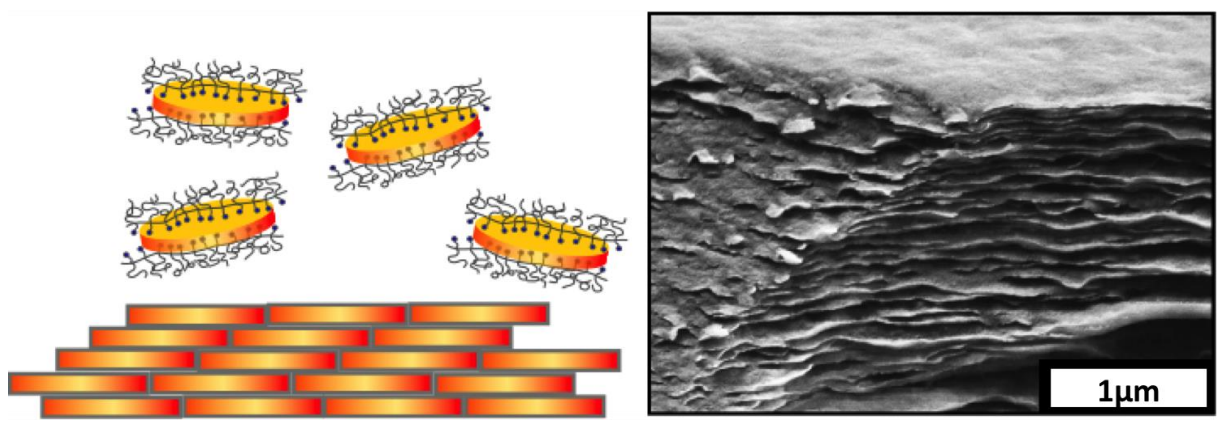

**Figure 7:** Schematic representation of a three dimensional layer-by-layer film formation similar to a mortar and brick formation with the PLL-PEG chains forming the mortar and the clay particles representing the bricks. Right is an SEM image taken from a 2.6μm thick layer, formed by spreading and drying an over-coated Laponite suspension. The sample was fractured to see the inside of the layer.

## Experimental

### Materials

The clay, Laponite XLG, used in our experiment was kindly provided by 'Rock Additives' and stored in a desiccator. The negative surface charge of Laponite crystal was 0.5-0.55nmol/g.[31] The comb-like polymer poly(L-lysine)-g-polyethylene glycol (short PLL-PEG from SuSoS Chemicals AG) consisted of a PLL(20kDa) backbone corresponding to about 100 lysine repeats to which PEG(5kDa) chains were grafted every g = 3.0 to 4.5 lysine repeats. 1mol PLL-PEG contains approximately 71.5mol free lysine, which corresponds to 71.5mol positive charges.[14,32] All materials were used as received.

### Preparation of clay-polymer nanocomposites

Polymer coated clay particles were obtained by first preparing a 1.0w% Laponite solution by adding 0.5g dry Laponite XLG powder to 49.5g deionized water. To ensure that the Laponite powder exfoliated completely the solution was vigorously stirred for 24 hours. After one day, it was filtered using a Millipore Syringe Filter from Sigma-Aldrich, with a pore size of 0.22µm. Simultaneously, a 1g/L PLL-PEG solution was prepared by dispersing 2mg polymer in 2mL deionized water, which was left to fully disperse for 24h as well. The amount of PLL-PEG chains per Laponite needed to balance the positive lysine charges and the negative charges on the flat clay surfaces was estimated to be 1.2mL of a 1g/L PLL-PEG solution per 50µL of a 1.0w% Laponite solution. Note that the clay dispersion was added slowly to the PLL-PEG solution in order to assure proper coating. This was done by step-wise addition of 5µL of 1w% Laponite solution to the 1.2mL of the original PLL-PEG solution.

### Imaging and particle sizing

PLL-PEG stabilized Laponite particles were imaged with a transmission electron microscope (TEM, TECNAI 20 from FEI of Thermo Fisher Scientific). A zeta-sizer Nano ZS (Malvern Instruments) was used in dynamic light scattering (DLS) mode to measure the size distribution of pure and PLL-PEG coated Laponite particles solutions; further their zeta potential was determined.

### Micro-rheology

The viscosity of the freshly prepared and aged samples with PLL-PEG coated samples were measured with a home-build optical tweezers setup build on an inverted Nikon Eclipse microscope, and analyzed with home-developed analysis methods.[19] A 1.2 µm large silica bead was inserted into the different clay suspensions introducing a change in the overall composition by less than 2%. A probe particle was weakly trapped in the optical tweezers beam (1064 nm) and its thermal motions monitored with a Sony IMX174 CMOS sensor CCD camera. The bead's traces in *x*- and *y*-direction were analyzed using the relation between the Laplace transform of the position-autocorrelation function and the generalized Stokes-Einstein relation.[19] Note, pure Laponite samples could not be analyzed after a year, as they were completely gelled.

**Preparation of liquid crystals**

In order to obtain a liquid crystal, a small drop of sample was taken and carefully placed onto a clean glass slide. The samples were then left to dry at room temperature in a closed room with little air-circulation, thus slowly increasing the clay-polymer composite concentration. The half-dried samples were observed with an optical microscope (OLYMPUS BX60) equipped with polarizer and analyzer. Images were taken in transmission mode using an AxioCam MRc5 (ZEISS) camera.

**Preparation of films**

Nanocomposite films were prepared by adopting a layer-by-layer method, reported by Tang et al.[33] For that we placed a small drop of clay-polymer hybrid solution on a clean glass substrate. A blade or spatula was used to spread the drop manually to form a very thin film. After allowing the film to dry for 2 hours, another drop of solution was spread on top of the first film, moving the blade in the same direction as the initial drop. This spreading-evaporation procedure was repeated five times. For the purpose of observing the cross-sectional structure of the final thicker films, the glass slide holding the firmly attached clay-polymer film was cut with a diamond cutter from the backside and then broken. The resulting fractured film was then imaged using a scanning electron microscope (SIGMA VP field emission SEM from Zeiss). Note such prepared films were kept in a desiccator until they were used for SEM imaging. This method provided us with highly reproducible results.

## Conclusions

In this article we present the first successful steric stabilization of Laponite particles in aqueous solutions that allows us to suppress any aging process. We achieved the steric stabilization by utilizing the attractive Coulomb interactions between the positive charges of polylysine chains and the negatively charged flat surfaces of the clay particles. The PEG chains attached to the polylysine backbone then stretched out into the water phase (a good solvent for PEG), thereby providing the steric stabilization. We demonstrate that gelation in all Laponite suspensions can thus be avoided and the viscosity of these suspensions maintains the same viscosity even after one year. It is this steric stabilization that allowed us to concentrate the samples through slow evaporation to reach the systems natural liquid crystal behaviour. We observed a columnar phase in the under- and perfectly-coated Laponite systems, while samples with excessive PLL-PEG in solution underwent a phase separation into PLL-PEG-rich droplets surrounded by a smectic, Laponite-rich phase. Finally, we showed that these sterically stabilized disk-like particles can be used to form thin transparent films with a 3D 'brick-and-mortar' structure, very similar to those in natural nacre (mother of pearl). Hence our system can be utilized as biomimetic materials with nanoscale structure.

## Acknowledgements

PX thanks his parents for financial support and EE thanks Daan Frenkel for fruitful discussions. YL acknowledges financial support from the Winton Program for Sustainable Energy.